\documentclass[12pt,journal, comsoc]{IEEEtran}
\usepackage[T1]{fontenc}
\usepackage{color}
\usepackage{subfigure}
\usepackage{amsthm}
\usepackage{amsmath}
\usepackage{amsopn}
\usepackage{tabularx}
\usepackage{booktabs}
\usepackage{xcolor}
\usepackage{graphicx}
\usepackage{framed}
\usepackage[inline]{enumitem}
\usepackage{eurosym}
\usepackage{setspace}
\usepackage{multirow}
\usepackage[style=ieee]{biblatex}
\usepackage{balance}
\usepackage{amsmath}
\usepackage[cmintegrals]{newtxmath}
\begin{document}
%
\title{\huge An End-to-End Integrated Computation and Communication Architecture for Goal-oriented Networking: A Perspective on Live Surveillance Video}
%
%
%
\author{Suvadip~Batabyal,
        Ozgur~Ercetin
\thanks{S. Batabyal is with the Department of Computer Science \& Information Systems at BITS Pilani, Hyderabad Campus, India. e-mail: sbatabyal@hyderabad.bits-pilani.ac.in}
\thanks{O. Ercetin is with the Faculty of Engineering and Natural Sciences, Sabanci University, Istanbul, TR. e-mail: oercetin@sabanciuniv.edu}  
}

\maketitle
\begin{abstract}
Real-time video surveillance has become a crucial technology for smart cities, made possible through the large-scale deployment of mobile and fixed video cameras. In this paper, we propose situation-aware streaming, for real-time identification of important events from live-feeds at the source rather than a cloud based analysis. For this, we first identify the frames containing a specific situation and assign them a high scale-of-importance (SI). The identification is made at the source using a tiny neural network (having a small number of hidden layers), which incurs a small computational resource, albeit at the cost of accuracy. The frames with a high SI value are then streamed with a certain required Signal-to-Noise-Ratio (SNR) to retain the frame quality, while the remaining ones are transmitted with a small SNR. The received frames are then analyzed using a deep neural network (with many hidden layers) to extract the situation accurately. We show that the proposed scheme is able to reduce the required power consumption of the transmitter by 38.5\% for 2160p (UHD) video, while achieving a classification accuracy of 97.5\%, for the given situation.

 
\end{abstract}



%

\section{Introduction}
\noindent The global market for smart city surveillance equipment – defined by Berg Insight as networked security technology to improve public safety levels in metropolitan areas – was worth \euro 6.5 billion in 2018 and is forecast to grow at a compound annual growth rate of 24.5\% to reach \euro 19.5 billion by 2023. A surveillance system has thus become a significant component of public safety, with numerous cameras (both planted and crowdsourced) capturing the environment. Live feeds from the cameras need to be analyzed in real-time to detect and track objects, such as vehicles breaking a traffic rule. However, such real-time assessment of subjective and semantic information in a video is a challenging task primarily for several reasons. First, the video needs to be analyzed at source to reduce the backhaul latency and streaming delay instead of cloud-based analysis. This requires the availability of significant computation capability at the source, which demands substantial power consumption. Second, continuous wireless multimedia streaming demands a constant network resource (in terms of bandwidth) and incurs continuous transmission power. However, video surveillance devices have a constrained power budget, especially when airborne; hence continuous streaming of high-quality surveillance is challenging. Third, such streaming is often performed under the cellular spectrum to improve spectral efficiency and reduce operational costs. This, in turn, may lead to the additional challenge of uplink interference with the nearby cellular base stations.

A streaming video, in general, is in  one of two states: \begin{enumerate*}[label=(\roman*)]
     \item \textit{steady state}, when the video is prosaic;
     \item \textit{transient state}, when the video has high situational content
 \end{enumerate*}.  While significant portions of these live streams are prosaic, intermittent incidents/events have a high situational significance and require the immediate attention of incident management unit(s) (IMUs) to ensure public safety. Streaming the prosaic portions of the live stream has less safety concern, so it is less of a problem if the received video is highly distorted or is primarily skipped. On the other hand, portions of the stream having high situational significance (may) impact public safety and require high quality and uninterrupted streaming. Several applications, such as the Wowza streaming engine developed by Wowza media systems, are used for streaming live surveillance video and can be integrated into any security product or service. \textbf{Therefore, surveillance systems need to be aware of the situation of the physical environment and prioritize streaming of video frames/segments, under constrained bandwidth, to capture important events promptly}.



\begin{figure*}
\includegraphics[width=1.0\linewidth]{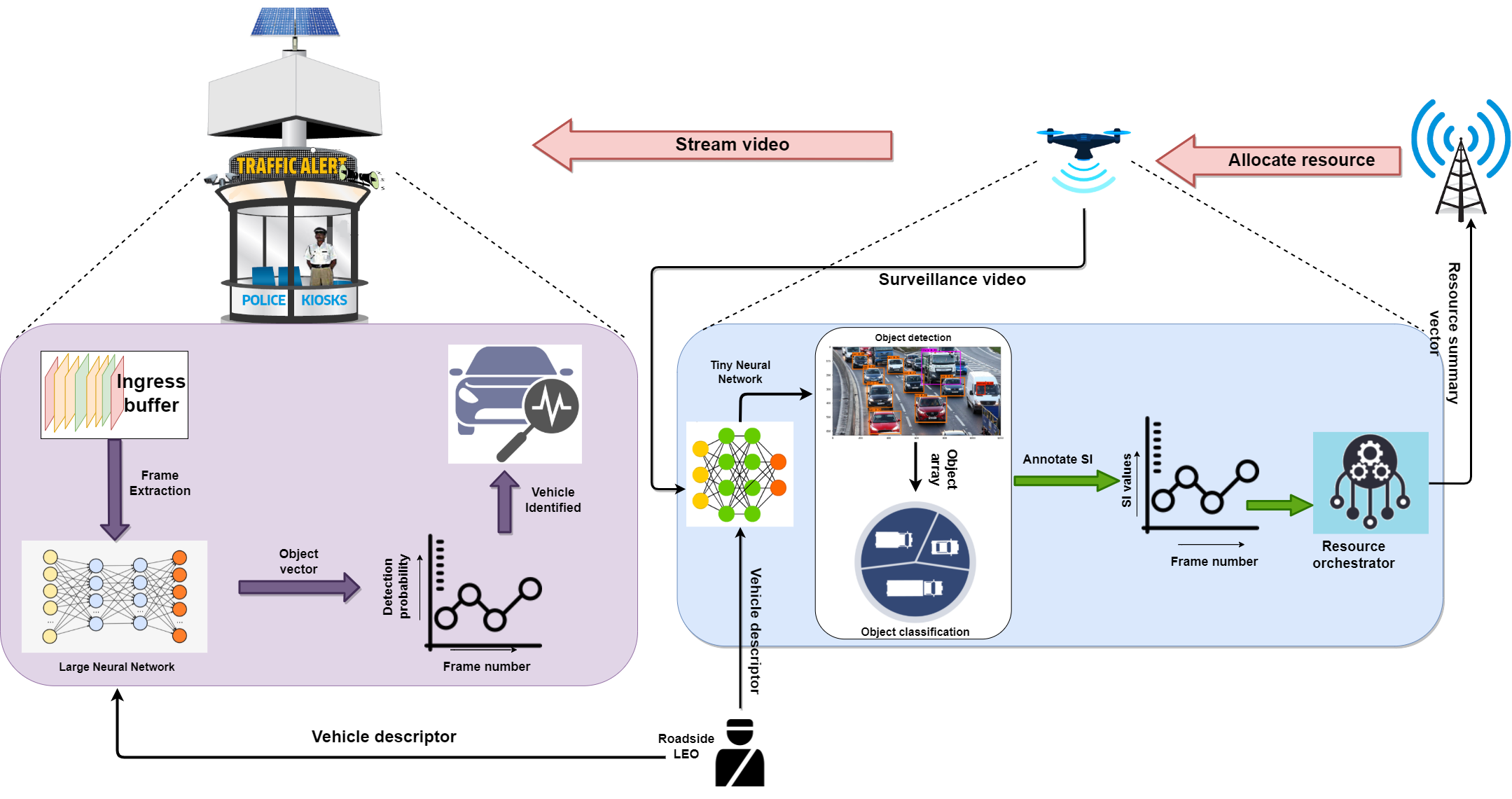}
\caption{System architecture}
    \label{fig:arch}
\end{figure*}

The situation-aware streaming addressed in this paper falls into the  emerging area of \textit{goal-oriented communication}  \cite{CALVANESESTRINATI2021107930}.  Unlike traditional communications, where the idea is to transmit all data reliably over an unreliable channel, \textbf{goal-oriented communication aims to deliver only the necessary information required to fulfill a given goal}.  Goal-oriented communication is gaining attention because of the widespread use of deep neural networks (DNNs). DNNs represent a potent computational tool that finds essential features in the data, and it is used in various fields, e.g., image and video processing.  In this paper, we propose an \textit{integrated computational and communication architecture}, where
a tiny neural network (TNN) is used by the source to identify the situational importance of video instances in real-time. The TNN, that is trained to identify a certain situation/event, can quickly identify a situation from the live-feed, albeit with a low accuracy. The frames, containing this situation, are then transmitted by the source with the required SNR, which are then analyzed at the destination with a DNN to accurately identify the situation. The proposed method has several advantages: First, the source incurs a small computational power, since it is using a tiny neural network. This also reduces the delivery latency, so the video is delivered almost in real-time. Second, since only the important frames are transmitted with a certain SNR, the transmission power incurred is small. This helps the source to efficiently allocate its limited resources to achieve the goal of the receiver. Third, the uplink interference is reduced, again because most of the time the video is in the steady state and is transmitted with a small power.

\section{Situation-Aware Streaming: Key Factors}
\subsection{Overview}
\noindent 
\textit{Situational awareness} is the perception and understanding of environmental factors and predicting future developmental trends under particular time and space conditions. The primary challenges of situation-aware streaming are: (i) the real-time identification of frames containing important events, (ii) optimal utilization of available bandwidth for transmission of such frames, and (iii) achieve a high Quality of Experience.


\subsection{Real-time identification of a Situation}
\noindent A situation (or scene) is defined as semantically related video frames with specific and similar contents. Scenes are primarily detected based on the contents of the frames. One may realize real-time identification of important events in video frames by real-time object detection and tracking; context modeling and semantic association; scene detection. While \textit{object detection} involves detecting a given object in a sequence of video frames, object tracking is the process of locating the motion of an object in a series of frames. \textit{Context modeling} refers to the object's identity, location, and scale, and it simplifies object discrimination by cutting down on the number of object categories. \textit{Semantic association} refers to associating meanings in formats that humans can understand along with the context. 

Identification of important scenes in a video had been extensively investigated.  For example, Gao \emph{et al.} used a deep learning approach, where a 3D ConvNet was used to learn the spatio-temporal features \cite{2}. The authors used the extracted video features to train a neural network (NN) which decides on the importance of a video frame to capture the \emph{interesting} scenes in real-time. Hu \emph{et al.} used a convolutional NN to extract emotional features 
from the frames \cite{Hu}. The authors  then trained a support vector machine (SVM) classification to confidently differentiate between video emotions. Finally, \cite{scene} provides a detailed survey of state-of-the-art techniques of scene detection in a video.

Situation-aware live streaming requires detecting specific objects/events in video frames and streaming them with high quality. However, identifying the frames having the desired object is computationally intensive and can result in streaming delays. There are a few studies comparing the processing speeds between different implementations of CNN/DNN with varying resolutions of video and image. For example, it has been shown that {\tt YOLOv4} and {\tt YOLOv3} have an average precision of 43.5\% and 34\%, respectively, with a processing speed of 65fps when run on Tesla v100. Similarly, \cite{CNN_comp} performed a comparison of SSD (single shot multibox detector), Faster RCNN (region-based convolutional neural network), and {\tt YOLOv3} on an Intel i5 8th generation processor on the COCO dataset. They showed that {\tt YOLOv3} is the fastest with a good average precision and therefore is most suitable for processing live-feed video. However, most studies concentrated on the performance of high-end systems, but did not consider low-end systems \cite{Foresti}.


\subsection{Resource aware transmission of live video}
Note that the source devices (such as roadside and airborne camera) are usually less capable compared to their fixed and application-specific counterparts.  In particular, the power and bandwidth allocation for such an airborne device, that may be the source or a relay for the video stream, should be optimized according to the resource limitations of the device. Multi-user video transmission was investigated extensively in the literature.  For example, \cite{lyapunov} used Lyapunov optimization to optimally allocate power and bandwidth such that the overall long-term average quality-of-experience (QoE) is maximized for all users. \cite{layered} investigated the power required to transmit a multi-layered video for optimal QoE, as measured in terms of video PSNR. \cite{UAV} investigated the problem of UAV relocation and adjustment of the transmission directionality, when the UAV's transmitted uplink signal  interferes with the transmissions in neighboring base stations (BSs). 
An uplink power allocation for UAV-assisted or UAV-sourced streaming is considered essential because (i) UAVs are battery powered and a large uplink power can be detrimental to their operational times;  (ii) uplink interference with the neighboring BSs should be avoided. 


\subsection{Quality of Experience}
\noindent Quality of experience (QoE) is defined as the \textit{user-perceived} quality of the multimedia service. 
QoE measurement and prediction for adaptive video streaming have been studied extensively in the literature. In addition, there are several commercial software available for use by cellular operators to understand the performance of their networks.
However, available QoE models often ignore the contextual importance of the video segments. 
In fact, with many diverse streaming applications, the goals of these applications become more critical in defining how QoE is measured \cite{10.1145/2160158.2160161}.  

In goal-oriented communication, the communication itself is not an end, but a means to achieve some goals of the communicating parties.  Hence, in this paper, we consider video streaming to perform real-time vehicular traffic monitoring. Our goal is to accurately identify the existence of a critical object with as little data transferred from the source to the receiver, and we define QoE as the accuracy of the detection.

\section{Situation-Aware Streaming: System Architecture}
\noindent We consider a real-time vehicular traffic monitoring system as a typical example of a situation-aware video streaming application. 
Figure \ref{fig:arch} shows the architecture of the proposed situation-aware streaming framework. A low-end monitoring device (e.g., an airborne vehicle such as a drone) monitors and captures the video of a certain area, and streams this video to a ground IMU inside a kiosk. 


\subsection{Architecture and Implementation}

\noindent We propose an integrated architecture of computation and communication to support  goal-oriented communication. In this architecture, application intentions are interpreted by the source and/or relay devices in the network.  Hence, each source and/or relay operates a \textit{surrogate} of the original application.  The surrogate application is an epitomized version of the original that can be run with higher speed and implemented with lower complexity. The surrogate application provides the source/relay the necessary association of data with the goal, which is used to optimize the networking operations such as routing, scheduling and transmission power control.

In the considered use case, the drone feeds the live-video to a tiny neural network (TNN) module on the device, consisting of only a few hidden layers, which can identify the desired vehicle \emph{quickly} albeit with a low accuracy. The TNN module consists of two components, viz., object identification and object classification. The TNN determines the probability of detection of each vehicle type in each frame. The roadside law-enforcement officer (LEO) feeds a vehicle description to the TNN. Based on this input, the TNN module annotates each frame with a scale-of-importance (SI) value, which denotes the probability that the given vehicle is present in a given frame. Finally, the resource orchestrator sends a resource summary vector (RSV) to the nearest radio access network (RAN). The RSV consists of maximum SNR required to transmit a given frame from the drone. Note that the resources are allocated for each video frame, and the RSV is sent for each video segment. We assume that the communication is performed under the cellular spectrum, and a traditional sum-rate maximization based scheme is used for allocating the uplink sub-carriers \cite{uplink}. As the live-feed is received at the kiosk, it is analyzed using a large neural network, using the vehicle descriptor provided by the LEO, to identify the designated vehicle with a high probability.

\subsection{Video Characteristics and Processing}
\noindent Live video stream is encoded as a Scalable Video Coding (SVC) H.264 video. The video is divided into segments or group of pictures (GOPs) typically of equal length and has a closed GOP structure with a frame rate of 30Hz. Each of these frames, depending on its size (in bytes) and maximum transmission unit (MTU) size, may be further divided into multiple slices. The video consists of I, P and B frames and one I-frame occurs every 30 frames. The GOP size of the video is a design parameter but in this work it was taken as 1 second. 

The video is first submitted to the TNN for processing. The object detection component of the TNN is run frame-by-frame to identify the objects that are present in each frame. For each frame, the object detection component creates an object array and passes it to the classification component. The classifier creates a 3D array representing the detection probability of a vehicle type and make in each frame, i.e., for each frame it assigns the prediction probability for each vehicle type and make. Moreover, it also predicts the number of unique objects present in the video. For example, if a given car (car type and make) is present in two consecutive frames, it is detected as a unique object and not different object. This is useful since there may be two cars of same type (and make) in different frames and only one of them may be an object of interest. Finally, based on the vehicle descriptor provided by the LEO, the frames with the designated vehicle with a certain threshold detection probability, are assigned an SI value of 1. The remaining frames are assigned an SI value of 0. This is called the frame vector. For example, if frame number 50 contains a given vehicle with a probability say 0.8, and if the threshold value is 0.8, then the frame is assigned an SI value 1; if the threshold value is greater than 0.8, then the frame is assigned a SI value of 0. 

\subsection{SI based Resource Orchestration}
\noindent An SI value of 1 denotes that the frame should be transmitted with the original quality and there is no desirable deterioration in frame quality. Due to dynamic wireless channel conditions, this implies that the sender should allocate a certain transmission power in the time-slot when the frame is to be sent to ensure a certain SNR. Note that the threshold that is selected to create the frame vector depends on the power budget of the node and the desired detection probability on the receiver side. However, we assume that: (i) the sender uses the cellular spectrum for transmission and, (ii) there are several other devices which are operating under the given RAN. Therefore, to minimize interference, the power allocation should be done under the control of the BS. Hence, the sender sends a resource summary vector (RSV) to the base-station. The RSV contains the required SNR for each frame. Based on the RSV, the BS assigns/reassigns the power for other devices in the network for the given slot and also sends confirmation to the sender.

\section{System Evaluation and Results}
\noindent The proposed architecture is evaluated using a H.264(SVC) encoded sample traffic video \cite{pexels}, which was given as input to three different {\tt YOLO} (You Only Look Once \cite{redmon2016look}) versions viz.,  {\tt YOLOv3(tiny)}, {\tt YOLOv2} and {\tt YOLOv3} for object (vehicle) detection. Note that {\tt YOLOv3(tiny)}, {\tt YOLOv2} and {\tt YOLOv3} have 13, 26 and 106 hidden layers, respectively. We encode the sample H.264 video in five different resolutions viz., 144p, 360p, 720p, 1080p and, 2160p. The object classification was done using two {\tt Resnet} versions viz., {\tt Resnet-18} and {\tt Resnet-50}. The object detection and classification was run on two different machines having different configurations in order to understand the delay distribution (in seconds/frame). Note that {\tt YOLO} is one of the object detection algorithms available and other alternatives may give better or worse performance depending on the video.  Nevertheless, the main idea and the approximate performance gains of the proposed architecture remain the same.

\subsection{Experimental setup}
\noindent The video stream between the source and receiver is emulated by using {\tt ffmpeg} software. The video was streamed from a system having lower configuration to a system with a higher configuration. The system with lower configuration may represent an aerial vehicle and is denoted as system-A; and the one with higher configuration may represent the kiosk and is denoted as system-B. System-A has a single-core Intel Celeron processor with a 4GB RAM, frequency of 1GHz, 4GB RAM, and integrated graphics with a frequency of 350MHz. System-B has i7-6700HQ processor, frequency of 2.60GHz, 8GB RAM and Nvidia GeForce GPU. Accordingly, the drone is running a {\tt YOLOv3(tiny)} model with {\tt Resnet-18} due to low system configuration; and the kiosk is running a {\tt YOLOv3} model with {\tt Resnet-50}. In order to emulate a power allocation algorithm leading to a specific SNR, we used the {\tt clumsy} software to introduce bit errors during transmissions. We assumed a simple QPSK modulation and an AWGN channel to find the desirable bit-errors for a given energy per bit to noise power spectral density ratio. Without affecting the generality, this approach may be used for other modulation and channel characteristics as well. We also assume that the transmissions are occurring in cellular frequency with a bandwidth of 20MHz.

\subsection{Results}

\subsubsection{Computation Delay}
\noindent Table \ref{tab:delay} shows the computation delay (in seconds per frame) for the video under different resolution on system-A and system-B (shown as system-A/system-B), respectively, without considering the SI values. The first value shows the computation delay on system-A and the second value on system-B. On system-A, significant time is spent for object classification compared to object detection. This can be observed from the difference in (average) computation delay for system-A between {\tt Resnet-50} and {\tt Resnet-18} for all  {\tt YOLO} versions and for all video qualities. The average delays for the {\tt YOLO} versions on system-A for {\tt Resnet-50} are 2.64s, 3.87s and 3.60s, respectively. Similarly, the average delays for all the {\tt YOLO} versions and for {\tt Resnet-18} on system-A are 1.25s, 1.29s and 1.89s, respectively. The difference in delay between two {\tt Resnet} versions is greater compared to the difference between any two {\tt YOLO} versions for system-A. Therefore, it can be inferred that for system-A, the object classification primarily contributes to the computation delay. Meanwhile, for system-B, the computation delay is primarily contributed by object detection. This supports the rationale for using {\tt YOLOv3 (tiny)} with {\tt Resnet-18} on a drone, and {\tt YOLOv3} with {\tt Resnet-50} on the kiosk.

\subsubsection{Detection Probability and SI}
\begin{figure*}
    \centering
\subfigure[]{\includegraphics[width=0.8\linewidth]{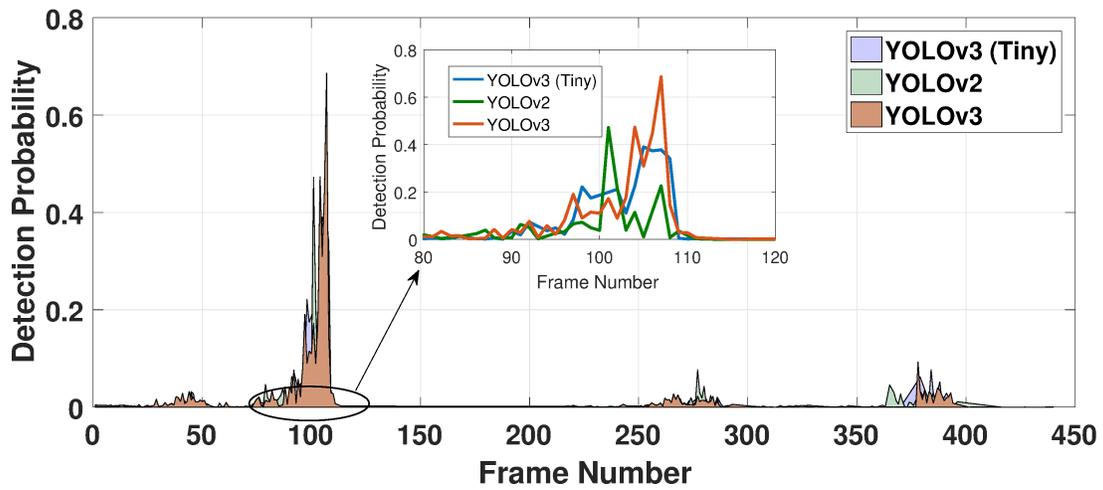} \label{fig:1a}}
\subfigure[]{\includegraphics[width=0.8\linewidth]{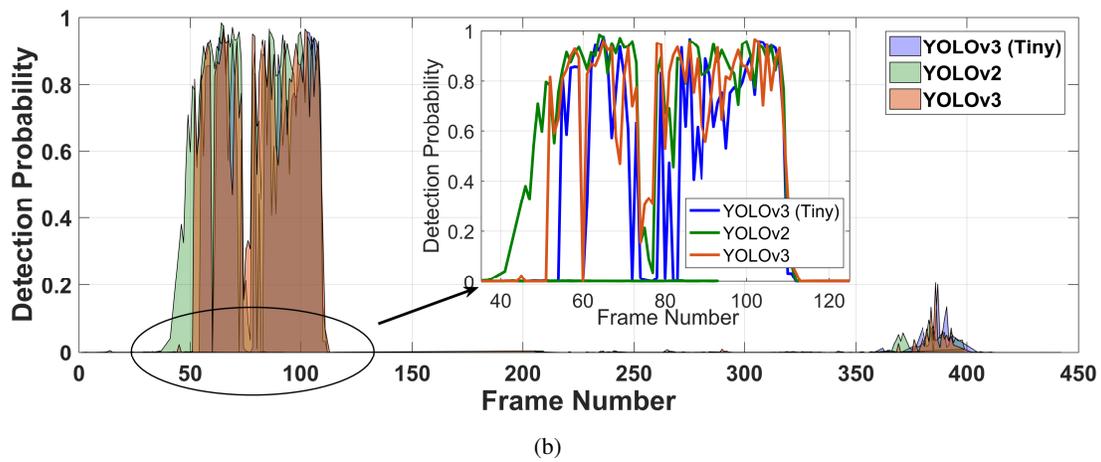}\label{fig:1b}}
\subfigure[]{\includegraphics[width=0.8\linewidth]{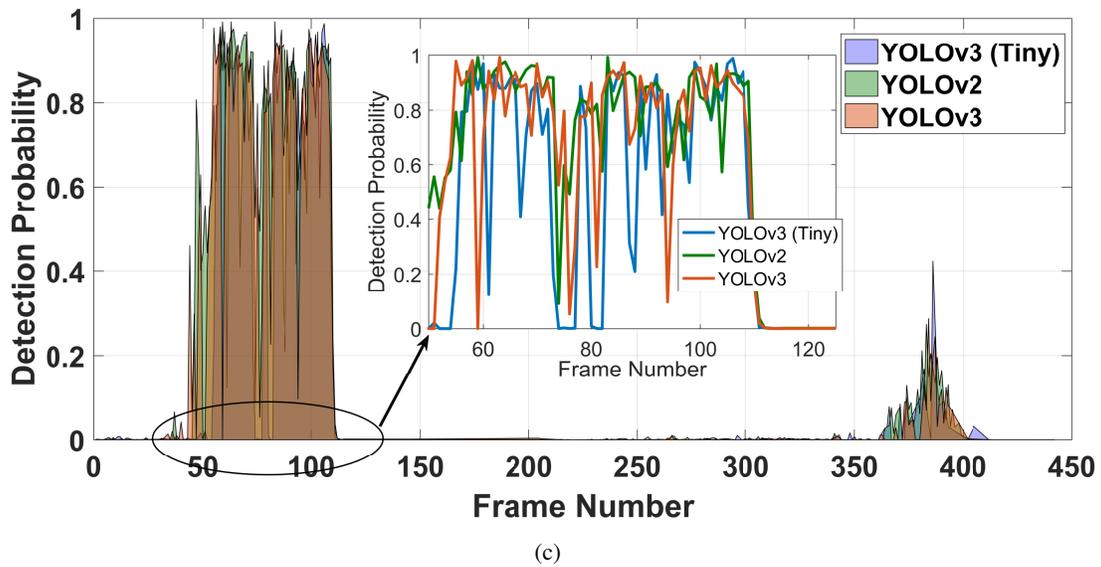}\label{fig:1c}}
    \caption{Detection probability of Ford Expedition 2017 in each frame for different YOLO versions for different video resolutions a) 144p  b) 720p  c) 2160p}
    \label{fig:detect_prob}
\end{figure*}

\noindent Next, we assign the SI values to the frames based on the detection probability and create the frame vector. For this we first log the detection probability of each vehicle make/model in each frame. The sample video contains 13 vehicles of different make. Figure \ref{fig:detect_prob} shows the detection probability of the vehicle Ford Expedition 2017 using different {\tt YOLO} versions and {\tt Resnet-18} for three different video resolutions viz., 144p, 780p, and 2160p.  The maximum detection probability (for the given vehicle) using {\tt YOLOv3(tiny)}  is only 0.4 for 144p video, whereas it is almost 1 for 720p and 2160p video. However, using {\tt YOLOv3}, the maximum detection probability (for Ford Expedition 2017) increases to about 0.65. Moreover, the number of frames where the vehicle is detected is small in 144p video compared to the video with higher resolutions. However, we can also observe, that for {\tt YOLOv3}, there are few frames towards end of the video with resolution 2160p, where the vehicle is detected with probability as high as 0.4. These are actually false detections, and therefore should be ignored.

In order to create the frame vector, a certain threshold value has to be selected based on the power budget of the sender. We have selected three threshold values viz., 0.4, 0.8 and 0.9. It is easy to see that the number of frames selected for threshold value greater than or equal to 0.9 will be a subset of the frames selected when the threshold value is greater than or equal to 0.4.

\subsubsection{Required Transmit SNR}
\begin{figure}
    \centering
    \includegraphics[width=1.0\linewidth]{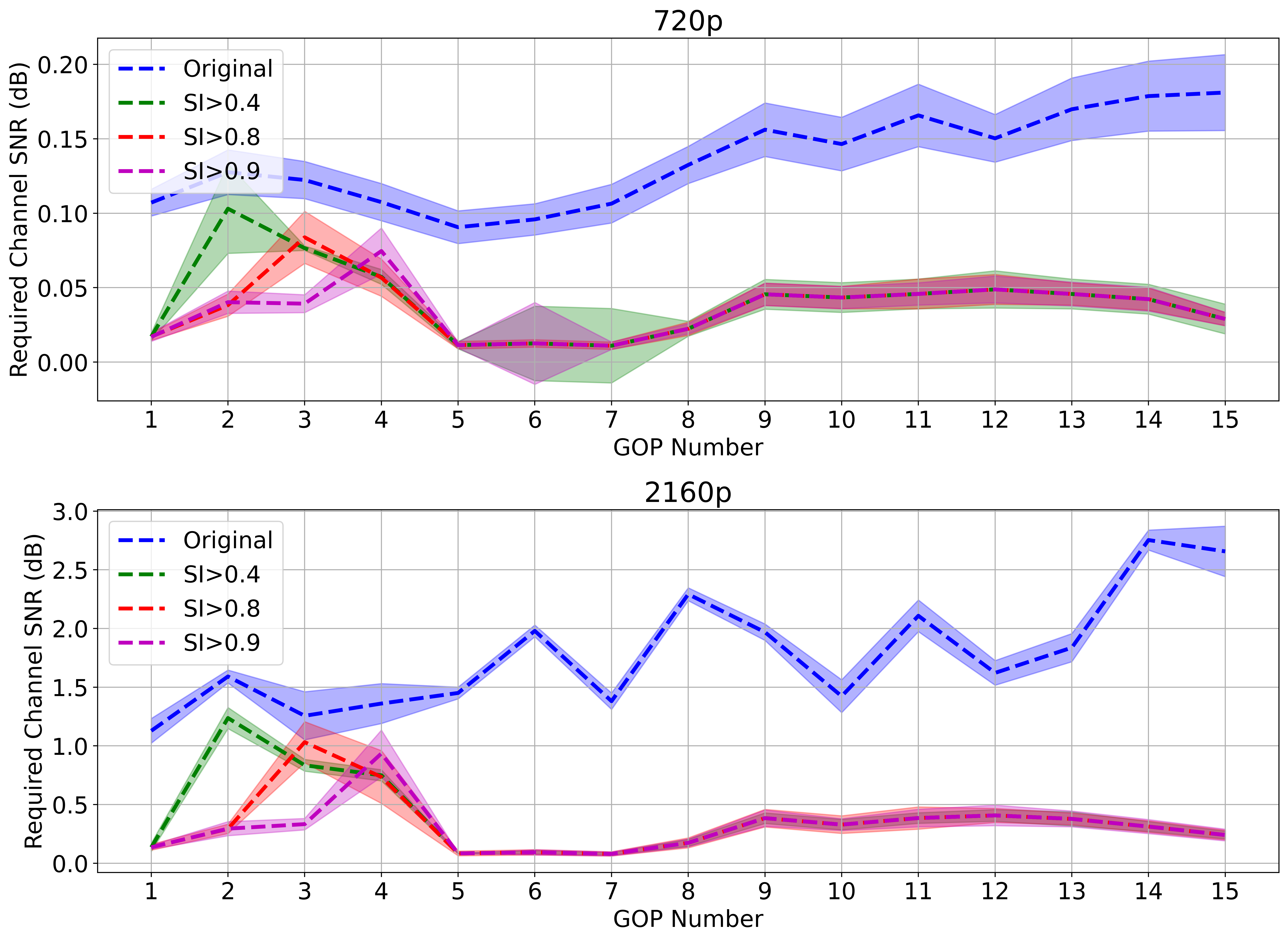}
    \caption{Figure showing the required SNR values with error for each GOP for resolution (a) 720p  (b) 2160p}
    \label{fig:my_label}
\end{figure}

\noindent The streaming of video with the desired SNR value was emulated by finding the approximate bit-error rate that would occur under the given network configurations (Section IV.A). This bit-error rate represents the expected error (and hence the degradation in video quality) that would occur when the sender streams the video with the actual allocated power from system-A to system-B. 
Figure \ref{fig:my_label} shows the required SNR to transmit the frames for two different video resolutions viz., the 720p and the 2160p video. The graphs show the required SNR for each GOP for different threshold values of SI. It can be observed that 2nd, 3rd, and 4th GOP have a higher requirement for SNR compared to other GOPs for both the types of video. Also, note that when SI$\geq$0.4, then the GOPs 2, 3, and 4 are transmitted with the highest SNR; similarly, for SI$\geq$0.8, GOP 2 and 3, and for SI$\geq$0.9 only GOP 3 is transmitted with a high SNR, compared to other frames for both the video types. Note that the size (in bytes) of GOP-2 > GOP-3 > GOP-4; so the required SNR is proportional to the size of the GOP.  The total noise normalized transmission power for 720p video without SI is $15.48$, and with SI$>0.8$ is $15.14$, which yields an improvement of $2.21\%$. For 2160p video, the improvement in transmission reaches approximately $38.5\%$. Our architecture is more advantageous if the source video has a high resolution. Note that with the prices of high-quality cameras decreasing rapidly, we expect that most surveillance cameras will have resolutions better than 720p (HD).  

\subsubsection{Classification Accuracy}
Recall that the goal is to identify a specific vehicle make and model in the live video. The received video is processed by System-B using a {\tt YOLOv3} model with {\tt Resnet-50}.   Table \ref{tab:detection} shows the detection probability of the vehicle Ford Expedition 2017 in the GOPs 1-3 averaged over 30 frames in every GOP. In the rest of the video, the detection probability of the specific vehicle is approximately zero. We note that the detection probability is highest when SI threshold is 0.4. This is because with a low SI most frames are considered important and transmitted with a high SNR.  However, this results in an excessive transmit power consumption.  Therefore, SI threshold should be chosen to minimize the transmission power while achieving a target detection probability.  In this example, choosing SI threshold as 0.8 appears as a good choice, since the detection probability is not significantly lower than that of 0.4, but uses overall $20\%$ less transmission power.

\begin{table*}[]
    \centering
    \caption{Computation delay (in seconds/frame) for system-A/system-B}
    \label{tab:delay}
    \small
    \begin{tabular}{|c|c|c|c|c|c|c|}
    \hline
        &  \multicolumn{3}{c|}{Resnet-50} & \multicolumn{3}{c|}{Resnet-18} \\
         \hline
         Video quality & {\tt YOLOv3(tiny)}  & {\tt YOLOv2} & {\tt YOLOv3} & {\tt YOLOv3(tiny)}  & {\tt YOLOv2} & {\tt YOLOv3}\\
        \hline
        144p & 2.22/0.04 & 3.36/1.42 & 2.96/1.81 & 0.96/0.04 & 1.06/0.50 & 1.41/0.64 \\
        \hline
        360p & 2.23/0.05 & 4.86/1.48 & 3.12/1.74 & 1.01/0.04 & 1.27/0.54 & 2.12/0.69\\
        \hline
        720p & 2.88/0.06 & 4.88/1.53 & 4.70/1.80 & 1.95/0.56 & 1.32/0.57 & 1.90/0.72\\
        \hline
        1080p & 2.30/0.09 & 3.67/1.53 & 3.14/1.82 & 1.12/0.08 & 1.24/0.58 & 2.02/0.75\\
        \hline
        2160p & 3.48/0.19 & 2.54/1.55 & 4.07/1.99 & 1.18/0.17 & 1.57/0.67 & 2.02/0.83\\
        \hline
        Average & 2.64/0.085 & 3.87/1.50 & 3.60/1.83 & 1.25/0.078 & 1.29/0.57 & 1.89/0.73\\
        \hline
    \end{tabular}
    
\end{table*}

\begin{table*}[]
    \centering
    \caption{Detection probabilities at the receiver}
    \label{tab:detection}
    \small
    \begin{tabular}{|c|c|c|c|c|c|c|c|c|c|}
         \hline
        &   \multicolumn{3}{c|}{2160p}& \multicolumn{3}{c|}{720p}\\\hline
        GOP number&{SI$\geq$0.4} & {SI$\geq$0.8}& {SI$\geq$0.9}&{SI$\geq$0.4} & {SI$\geq$0.8}& {SI$\geq$0.9} \\\hline
        1& 0.321	&0.4	&0.385	&0.26&	0.11	&0.13
 \\\hline
        2& 0.9921&	0.975&	0.964&	0.91&	0.89&	0.053
\\\hline
        3& 0.742&	0.738&	0.75&	0.74&	0.73&	0.67
\\\hline
    \end{tabular}
\end{table*}

\section{Open Challenges and Future Directions}
\noindent Future wireless networks need to support the delivery of only the {\em most informative} data for end-users in terms of how the data is timely, useful, and valuable for achieving their goals. In this aspect, traditional Key Performance Indicators (KPIs) are insufficient to understand the QoE of applications. Qualitative attributes of the applications should be identified based on possibly limited user/application feedback. In this work, we argue that an integrated computation and communication architecture is needed to effectively use the network resources while satisfying different goals of the applications.

Adding goal-oriented annotations to transmitted packets opens up many other possibilities such as using this information for routing, scheduling and spectrum allocation.  However, the efficiency of the solutions will be affected by the privacy concerns of the end-users, e.g., surrogate application should be designed according to the privacy requirements.  Additionally, surrogate application may evolve based on limited and incomplete feedback information from the original application.  Contemporary unsupervised ML methods such as reinforcement learning (RL), or meta-learning may help in this aspect.

\section{Conclusion}
\noindent In this work, we have proposed an end-to-end integrated computation and communication architecture for goal-oriented communication, where the objective is to selectively transmit live-feed instances from surveillance cameras based on a situation. This reduces the overall required transmit power, and delay while achieving the desired quality-of-experience. We show that the proposed scheme is able to reduce the required power consumption by 2.21\% and 38.5\% for 720p and 2160p, respectively, while achieving a classification accuracy of 89\% and 97.5\%, for the given situation.

\renewcommand*{\bibfont}{\normalfont\small}
\printbibliography
\balance
\end{document}